\documentclass[journal]{IEEEtran}
\usepackage{cite}
\usepackage{amsmath,amssymb,amsfonts}
\usepackage{subcaption}
\usepackage{textcomp}
\usepackage{hyperref}
\usepackage{nicefrac}
\usepackage{soul}
\usepackage{pifont}
\usepackage{hyperref}
\usepackage{xcolor}
\usepackage{cuted}
\usepackage[pdftex]{graphicx}
\usepackage{algorithm}
\usepackage{algpseudocode}
\usepackage{mathtools}
\usepackage{multirow}
\usepackage{epstopdf}

\begin{document}

\title{Deep-Learning Based Blind Recognition of Channel Code Parameters over Candidate Sets under AWGN and Multi-Path Fading Conditions}

\author{Sepehr~Dehdashtian,~Matin~Hashemi,~and~Saber~Salehkaleybar%
	
	{\color{blue} 
		\begin{flushleft}
			\footnotesize 
			This article is published. Please cite as S. Dehdashtian, M. Hashemi, S. Salehkaleybar, ``Deep-Learning Based Blind Recognition of Channel Code Parameters over Candidate Sets under AWGN and Multi-Path Fading Conditions," IEEE Wireless Communications Letters.
	\end{flushleft} }   
	
\thanks{The authors are with Department of Electrical Engineering, Sharif University of Technology, Tehran, Iran.  E-mails: dehdashtian.sepehr@ee.sharif.edu, matin@sharif.edu (corresponding author), saleh@sharif.edu.}
\thanks{Manuscript received Sep. 14, 2020; revised Dec. 13, 2020 and Jan. 22, 2021; accepted Jan. 30, 2021.}
\thanks{Digital Object Identifier XX.XXXX/WCL.2021.XXXXXXX}
}

\maketitle
\begin{abstract} 
We consider the problem of recovering channel code parameters over a candidate set by merely analyzing the received encoded signals. We propose a deep learning-based solution that I) is capable of identifying the channel code parameters for several coding scheme (such as LDPC, Convolutional, Turbo, and Polar codes), II) is robust against channel impairments like multi-path fading, III) does not require any previous knowledge or estimation of channel state or signal-to-noise ratio (SNR), and IV) outperforms related works in terms of probability of detecting the correct code parameters. 
\end{abstract}

\IEEEpeerreviewmaketitle


\section{Introduction}
\label{Introduction}

Forward error correcting (FEC) codes have been utilized to improve the performance of communication systems by detecting or correcting the errors occurred through noisy channels. As more users join wireless networks, allocating communication resources like channel bandwidth become more challenging. 
One of the main approaches to address the challenges in utilizing communication resources is to replace fixed transmission parameters with adaptive modulation and coding (AMC) in which the transmitter changes the modulation scheme and coding rate frequently to adapt to changing channel status. To achieve this goal, the receiver needs to be informed about transmission parameters with a signaling protocol whenever the transmitter adjusts these parameters. However, the main drawback of this approach is that the signaling messages occupy part of the channel bandwidth. 
To resolve this issue, blind recognition algorithms have been proposed to \emph{recover} the transmission parameters without any signaling from the transmitter. In this scenario, the receiver tries to identify the parameters merely from the received encoded signal (Fig.~\ref{fig:transceiver_model}). 
The blind recognition algorithms have several applications in cognitive radio and wireless sensor networks.

Previous approaches in blind recognition of channel codes fall in two main categories. In the first category, blind recognition algorithms aim at estimating unknown numerical parameters. For instance, code-word length $n$ and information length $k$ are estimated for convolutional codes \cite{marazin2011} and LDPC codes \cite{bonvard2018}. 
As another example, interleaver parameters are estimated for convolutional codes \cite{xu2019} and Reed-Solomon codes \cite{swaminathan2018}.

The algorithms in the second category, however, recover the parameter(s) only among a \emph{candidate set}. This is because many standard AMC schemes do not freely adjust all possible parameters, and instead, only select among a set of pre-defined parameters \cite{shen2019, wang2020, ni2020, xia2014, moosavi2014, yu2016, condo2017, wu2018, wu2019, sun2019, qin2019, jalali2020, condo2018, swaminathan2017(2), xia2014(2),liu2018,liu2020, wang2020nested}. 
Our solution falls in this category, i.e., recovers the code parameters among a candidate set.

In \cite{shen2019, wang2020, ni2020}, the candidate set consists of different coding schemes. 
For instance, the blind recognition algorithm proposed in \cite{ni2020} recovers the selected coding scheme among QC-LDPC, SC-LDPC, and convolutional codes. 
In \cite{xia2014, moosavi2014, yu2016, condo2017, wu2018, wu2019, sun2019, qin2019, jalali2020, condo2018, swaminathan2017(2), xia2014(2),liu2018,liu2020}, however, the coding scheme is fixed and known, and the candidate set consists of different parameters for that coding scheme. This type is the focus of our work and we try to recover the code parameters for a specific coding scheme such as the code rate. 
For instance, IEEE 802.11n WiFi standard with code-word length $n=648$, employs only four different LDPC codes with code rates $\{1/2, 2/3, 3/4, 5/6\}$ \cite{xia2014}.

Most previous methods are limited to AWGN channels \cite{xia2014, moosavi2014, yu2016, condo2017, wu2018, wu2019, sun2019, qin2019, jalali2020, condo2018, swaminathan2017(2)}. They mostly rely on theoretical analysis and signal processing algorithms based on characteristics of AWGN channels. 
In most real applications, however, channel impairments such as fading and multi-paths effects need to be considered in devising a resilient solution, which often break the employed AWGN-based mathematical analysis.

On the other hand, deep-learning based approaches have recently been proposed as well \cite{qin2019, shen2019, wang2020, ni2020}. Compared to classical methods, deep learning provides the opportunity to be more accurate, and also, more resilient to signal conditions. However, effective training of deep neural networks for blind code recognition is challenging, because the number of possible code-words to be seen in the training phase grows exponentially with $n$, i.e., code-word length.

This paper combines classical signal processing with deep learning in order to significantly reduce the number of samples needed to effectively train the neural network, and as a result, to be able to achieve very accurate blind recognition over different signal conditions. In specific, the received encoded signal is first processed and then fed into neural network (Fig.~\ref{fig:model}).  

The proposed approach has the following advantages over previous works: 
I) It works for different coding schemes, and is not tied to any specific coding.  
II) It also works under any channel impairment settings, including AWGN, fading, or multi-path conditions. 
III) No previous knowledge of channel state or signal-to-noise ratio (SNR), or their estimation is required in our method.
IV) Experiment results show that the achieved accuracy is much higher compared to previous methods, across different coding schemes and signal conditions. 



\section{Preliminaries}
\label{sec:TrModel}

\subsection{Basic Transceiver Model}

\begin{figure}[tp]
	\centering
	\includegraphics[width=0.95\columnwidth]{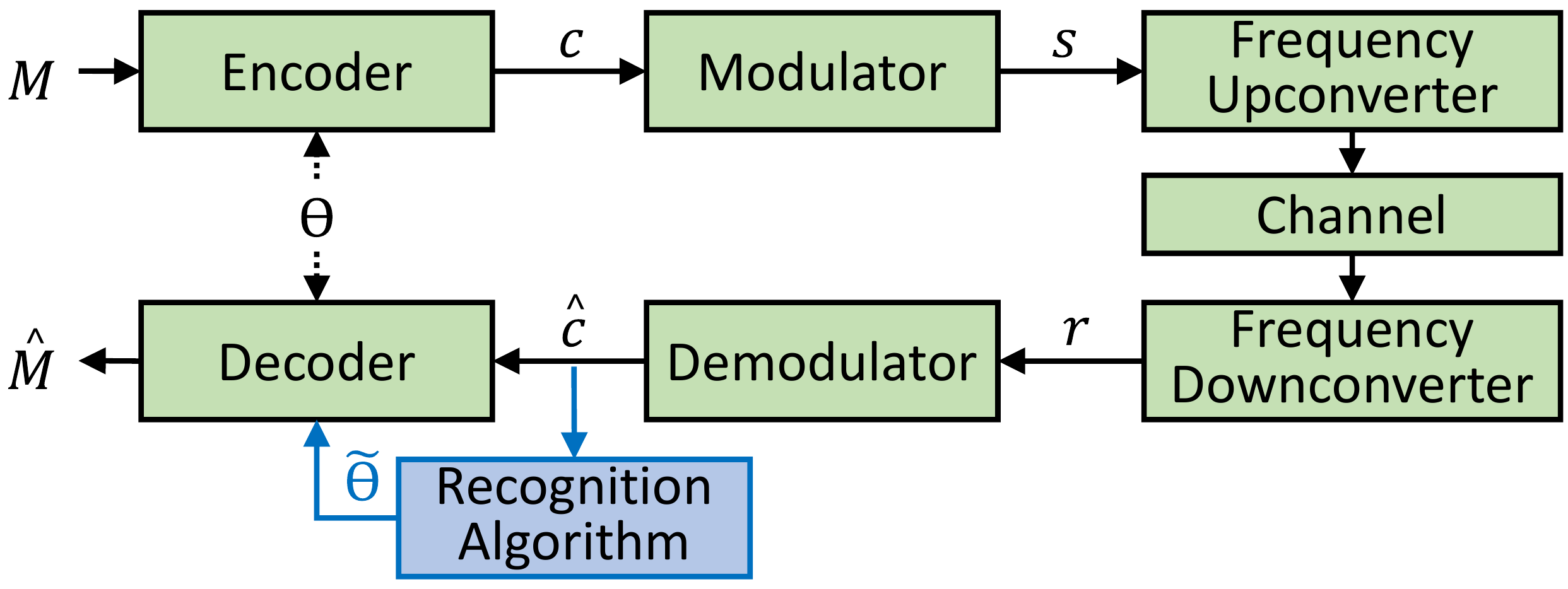}
	\caption{
		Green color: block diagram of a basic digital wireless transceiver model. 
		Blue color: An additional block which automatically recognizes the channel code parameter. 
	}
	\label{fig:transceiver_model}
\end{figure}

The green blocks in Fig.~\ref{fig:transceiver_model} show a basic digital wireless transceiver model. 
At the transmitter side, message $M$ with $K$ bits is first encoded into codeword $c$ with $n$ bits ($K<n$). Code rate $R$ is defined as the ratio $\frac{K}{n}$. Both $M$ and $c$ are in Galois field of two, i.e., GF(2). Codeword $c$ is formed based on generator matrix $G$ as the following:
\begin{equation}
\label{eq:c}
c_{1 \times n} =  M_{1 \times K} ~.~ G_{K \times n}.
\end{equation}

Next, vector $c$ is modulated and converted to analog domain. Then, frequency up-conversion is performed, and the generated signal is transmitted through the wireless channel. 

At the receiver side, after frequency down-conversion and demodulation, codeword $\hat{c}$ is obtained. Due to non-idealities that exist in  channel, transmitter, and receiver, $\hat{c}$ is not necessarily equal to $c$. 
Note that the demodulator provides every bit $i$ in vector $\hat{c}$ as a log-likelihood ratio (LLR) defined as 
\begin{equation}
\label{ell}
\log{\big(P\{\hat{c}[i]=0\} / P\{\hat{c}[i]=1\}\big)}
\end{equation}
where $P\{\hat{c}[i]=0\}$ and $P\{\hat{c}[i]=1\}$ denote the probability of bit $\hat{c}[i]$ taking value $0$ and $1$, respectively. 
The provided LLRs are processed by the decoder block to obtain message $\hat{M}$. 

\subsection{Problem Definition: Blind Channel Code Recognition}
\label{sec:problem}

There exist different parameters for a coding scheme. For instance, IEEE 802.11n WiFi standard employs four different LDPC codes with different code rates for code-word length $n=648$. 
Normally, the parameters are adapted according to the channel conditions and desired data rates. 
Without blind recognition (i.e., without the blue box in Fig.~\ref{fig:transceiver_model}), the selected code parameters (shown as vector $\theta$ in Fig.~\ref{fig:transceiver_model}) need to be sent to the receiver side as well. Code parameter vector $\theta$ is required by the decoder in order to correctly decode the message. 

However, to save part of the bandwidth and to provide the possibility of continuous modifications, the overhead of this negotiation can be removed by having the receiver obtains the selected parameters (shown as $\tilde{\theta}$ in Fig.~\ref{fig:transceiver_model}). This can be achieved by employing a blind channel code recognition algorithm. 
The  algorithm analyzes the received data and automatically selects $\tilde{\theta}$ from a set of known candidates. For instance, in IEEE 802.11n WiFi standard with $n=648$, the candidate set for $\theta$ is the code rates $\{1/2, 2/3, 3/4, 5/6\}$.

Similar to \cite{xia2014, moosavi2014, yu2016, condo2017, wu2018, wu2019, sun2019, qin2019, jalali2020, condo2018, swaminathan2017(2), xia2014(2),liu2018,liu2020}, we consider that the code parameters inside the candidate set are known, i.e., the set of corresponding generating matrices is known. 
It is noteworthy to mention that the problem addressed here is different from recognition of the coding scheme itself (e.g., recognition between LDPC and convolutional) as in \cite{shen2019, wang2020, ni2020}.

\begin{figure}[tp]
	\centering
	\includegraphics[width=\columnwidth]{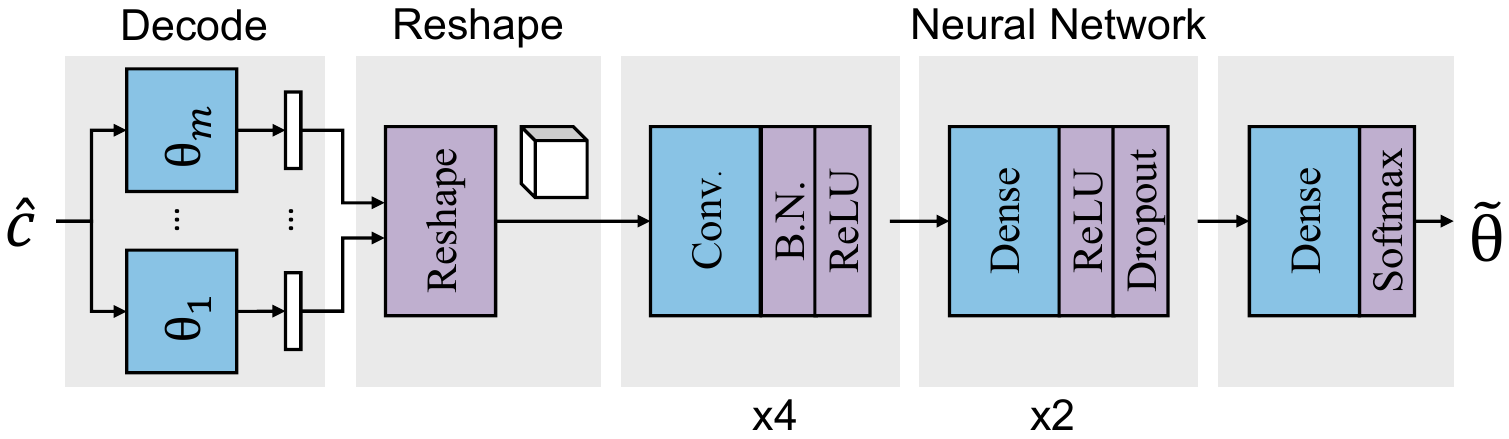}
	\caption{Proposed blind channel code recognition method, composed of decode, reshape, and neural network stages.}
	\label{fig:model}
\end{figure} 


\section{Proposed Recognition Algorithm}
\label{sec:alg}

\subsection{Deep-Learning Based Method}
\label{sec:alg:model}

The proposed solution combines classical signal processing with deep learning. See Fig.~\ref{fig:model}. In specific, the received encoded signal is first processed in order to extract a number of features, and then, the extracted features are combined and fed into deep neural network. 

\vskip 2mm
\emph{Rationale:} 
Intuitively speaking, the additional processing which is performed in the beginning helps the neural network to better capture different patterns. This is because now the network is provided with processed (as opposed to raw) data. 
As a result, neural network training becomes simpler and more effective. 
In specific, the number of samples required to be seen in the training phase is reduced significantly. This also allows to reduce the network size, and hence, the computational requirements. In addition, more effective training leads to higher accuracy across various signal conditions.

\vskip 2mm
\emph{Feature Extraction:} 
As shown in Fig.~\ref{fig:model}, in order to extract features, $\hat{c}$ is processed by decoding blocks with parameters $\theta_1$, $\theta_2$, $\cdots$ $\theta_m$. In other words, we decode $\hat{c}$ for a known coding scheme with parameter $\theta_j$ for $j=1,\cdots,m$. Every block represents one of the candidate parameters in the candidate set. Generated output vector from block $\theta_j$ is of size $n$.

In \cite{yu2016}, the average value for every output vector is computed, and the one with the maximum average is considered as the parameter $\tilde{\theta}$. This method is only accurate for high SNR ranges in AWGN channel. For lower SNR ranges or multi-path fading channels, these output vectors need to be processed by more complex methods as discussed below in order to accurately predict $\tilde{\theta}$.

\vskip 2mm
\emph{Reshape:} 
In our proposed solution, every output vector is folded into a two-dimensional matrix of size $\sqrt{n}$. As shown in Fig.~\ref{fig:model}, all such reshaped 2D matrices are stacked to form a 3D feature matrix of size $\sqrt{n} \times \sqrt{n} \times m$.  
Next, the extracted 3D feature matrix is processed by the following  network.

\vskip 2mm
\emph{Neural Network:}
The neural network consists of four convolutional layers. Convolutional kernel sizes are $1\times1$ in the first layer and $2\times2$ in the next layers. 
The employed activation function is ReLU. 
Batch normalization \cite{batch_norm} is added after every convolution layer in order to reduce internal covariate shift and improve the performance.

Next, the output of the convolutional layers are fed into dense neural network layers in order to classify the code parameter $\tilde{\theta}$. 
In specific, this network consists of three dense layers along with ReLU activation functions and softmax. 
Dropout \cite{dropout} is added after the non-linearity in order to apply regularization and avoid overfitting.

The output size of the last dense layer is $m$, i.e., the number of candidates. 
We employ cross-entropy (CE) as the loss function. The proposed model is trained using back propagation. The employed optimization algorithm is Nadam. 

\vskip 2mm
\emph{Complexity Analysis:}
Computation complexity of classifying an input signal with the proposed solution is in the order of $O(nm)$ where $n$ is the code-word length and $m$ is the number of candidates. 
Note that the proposed method has lower computational complexity compared with other deep-learning based methods. 
For instance, for the LDPC code in Section \ref{sec:exp:1}, our method requires about $115$M multiplications, while the method in \cite{wang2020} needs about $3.7$B multiplications.

\subsection{Channel Models}
\label{sec:alg:channels}

Many previous works rely on analytical techniques which are often limited to the AWGN channel model \cite{xia2014, moosavi2014, yu2016, condo2017, wu2018, wu2019, sun2019, qin2019, jalali2020, condo2018, swaminathan2017(2)}. The proposed deep-learning based solution, however, is capable of learning different hidden patterns in the signals and hence can support different channel conditions. We consider three channel models: 

Model 1: AWGN channel adds zero mean Gaussian noise with $\sigma^2$ variance to the signal. The signal to noise ratio (SNR) is given by $a^2 / \sigma^2$, where $a$ is the signal's amplitude. 

Model 2: Single-path Rayleigh channel models fading effects. Similar to \cite{xia2014(2)}, we consider Doppler frequency to gain ratio $f_D / f_s = 0.001$, where $f_D$ is the Doppler frequency and $f_s$ is the sample rate.

Model 3: The last channel model is dual-path Rayleigh channel that models the effects of both fading and multiple paths. We consider two paths and the same Doppler frequency as the previous channel model.

Note that one of the main advantages of our approach is that it does not require channel parameters such as signal-to-noise ratio (SNR) or channel state. This is in contrast to some other approaches that assume such parameters \cite{wu2019, yu2016, condo2017, liu2018} or estimate them with another algorithm \cite{moosavi2014, xia2014, xia2014(2), wu2018}.

\subsection{Training Method and Hyper-Parameter Selection}
\label{sec:alg:train}

Our dataset consists of 100,000 codewords for every code parameter and every SNR value.
The dataset is split in three portions, namely, train ($60\%$), validation ($20\%$), and test ($20\%$). Training is performed for multiple epochs. In every epoch, train dataset is randomly shuffled and provided to the model. Training phase terminates when the loss function on validation dataset stops changing for $10$ consecutive epochs.

There are a number of guidelines for selecting the hyper-parameters in neural network based algorithms. 
Here, for a certain set of hyper-parameters, the model is trained, and then, classification accuracy over the validation dataset is recorded. The procedure is repeated for different sets of hyper-parameters, and finally, the set of hyper-parameters that yields the highest recorded accuracy is selected.

We perform a grid search on the range of $3 - 5$ for the number of convolution layers, $64 - 256$ for the number of kernels in every convolution layer, $1 - 3$ for convolution kernel sizes, $relu$ and $tanh$ for the activation functions, and $1-4$ for pooling window sizes.  
We also perform a grid search on the range of $1 - 3$ for the number of dense layers, $512 - 2048$ for the number of their output neurons, and $relu$ and $tanh$ for activation functions.

We found that the following hyper-parameter selections achieve the best results over the validation dataset. 
The number of convolution layers is selected to be equal to $4$, and the number of kernels $128$. In the first layer, the kernel size is $1$, and in the other layers $2$. The selected activation function is $relu$, and the selected pooling window size is $1$, i.e., no pooling. 
The number of dense layers is selected to be equal to $3$. In the first two dense layers, $1024$ hidden neurons are used. The output dimension of the third dense layer is $m$, i.e., the number of code parameters in the candidate set. The selected activation function is $relu$. Dropout rate is $0.65$.

\begin{figure*}[tp]
	\centering
	\includegraphics[width=\textwidth]{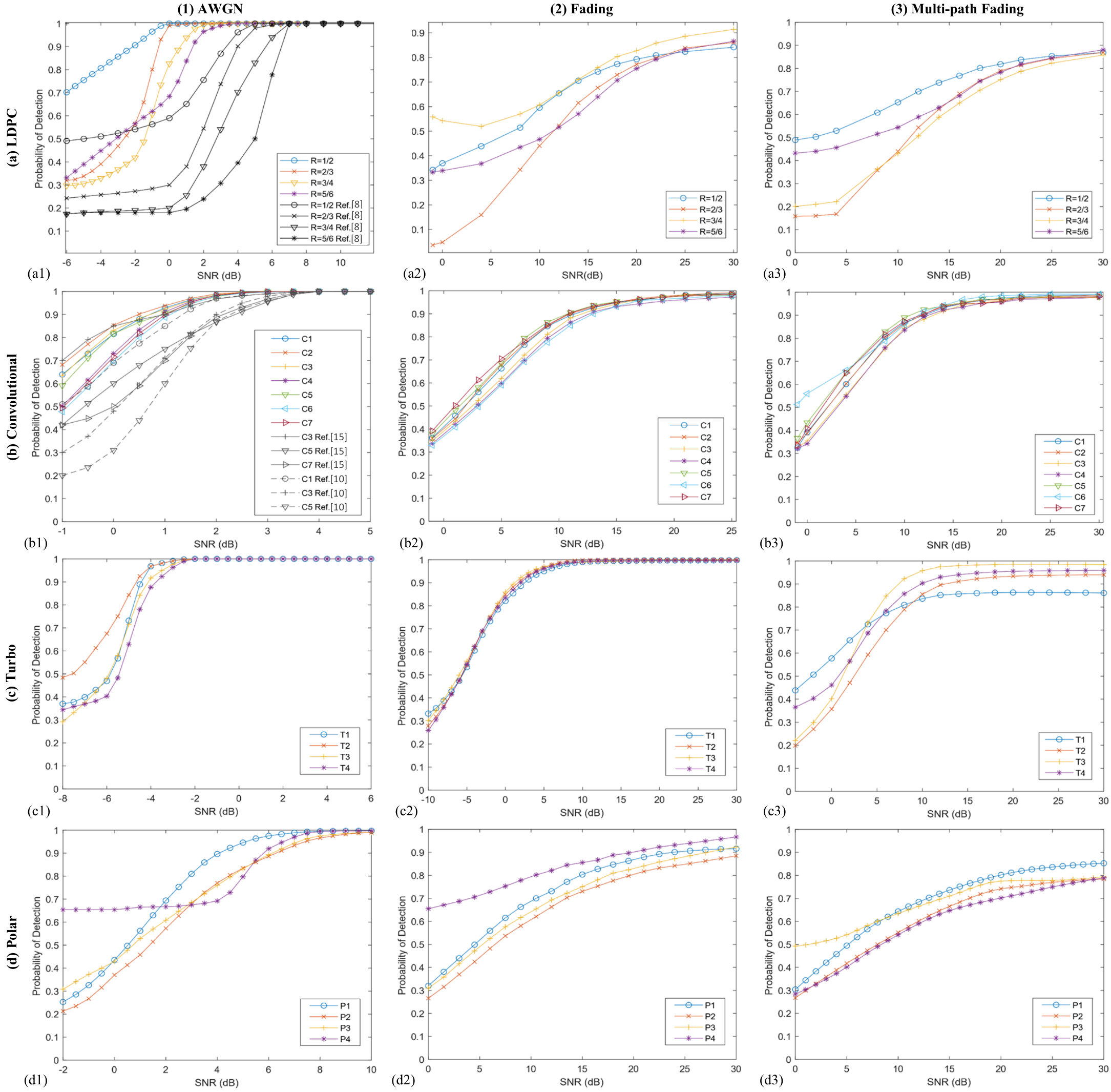} 
	\caption{Probability of detection among different (a) LDPC codes, (b) convolutional codes, (c) turbo codes, and (d) polar codes, versus different SNR ranges, for AWGN channel (1st column), Rayleigh fading channel (2nd column), and multi-path Rayleigh fading channel (3rd column). The black curves show the results for previous works, i.e., \cite{xia2014}, \cite{yu2016} and \cite{qin2019}. }
	\label{fig:exp}
\end{figure*}


\section{Experimental Results}
\label{sec:exp}

We experiment with four different channel codes, namely, LDPC, convolutional, turbo convolutional, and polar. Note that the proposed solution is not limited to any specific coding scheme. 
In every coding scheme, we consider the three channel models discussed in Section~\ref{sec:alg:channels}, along with BPSK  modulation. 
To evaluate the proposed solution and compare with previous works, \emph{probability of detection} is defined as 
\begin{equation}
\text{Probability of detection} = \text{N}_\text{true} / \text{N}_\text{total}
\end{equation}
for every code parameter in a candidate set, where $N_{true}$ is the number of truly-classified samples, and $N_{total}$ is the total number of samples in that specific class. This is the same metric used in \cite{xia2014, xia2014(2), yu2016}. 
Source code of the proposed solution is available online \cite{SourceCode}.

\subsection{LDPC Codes}
\label{sec:exp:1}

IEEE 802.11n (WiFi) standard with $n = 648$ is used to experiment with LDPC codes. The candidate set consists of four different LDPC codes with code rates $R = K/n = 1/2$, $2/3$, $3/4$, and $5/6$. This is the exact same candidate set as the blind recognition method in \cite{yu2016}.
Belief Propagation (BP) algorithm is employed to decode LDPC codes.

Fig.~\ref{fig:exp}(a) presents the probability of detection among the LDPC codes, versus different SNR ranges, for the three channel models discussed in Section~\ref{sec:alg:channels}, namely, AWGN, single-path Rayleigh fading, and dual-path Rayleigh fading channels with delay spread $10\mu$s. 

In addition, Fig.~\ref{fig:exp}(a1) compares our results with the probability of detection in the blind channel code recognition method in \cite{xia2014} under AWGN channel.
As the figure shows, the proposed solution achieves more accurate results compared to previous works. For example, at $0$~dB SNR, the probability of detection is between $50\%$ to $75\%$ higher compared to \cite{xia2014}. 
The proposed method reaches about $100\%$ accuracy at about $2$~dB SNR while the method in \cite{xia2014} reaches this point at about $7$~dB SNR. 
In addition, the probability of detection in all cases reaches above $60\%$ after $0$~dB SNR, and after $6$~dB SNR, in the proposed solution and in \cite{xia2014}, respectively.

\subsection{Convolutional Codes}
\label{sec:exp:2}

The candidate set for convolutional codes consists of code rate $1/2$ with seven different generator polynomials (in octal representation) as the following: $C1=(5,7)$, $C2=(15,17)$, $C3=(23,35)$, $C4=(53,75)$, $C5=(133,171)$, $C6=(247,371)$, and  $C7=(561,753)$, and constraint length $3$, $4$, $5$, $6$, $7$, $8$, and $9$, respectively. The information length $K$ is set to $50$. 
This is the same candidate set employed in the blind channel code recognition methods in \cite{moosavi2014}, \cite{yu2016}, and \cite{qin2019}. Viterbi algorithm is employed to decode convolutional codes.

Fig.~\ref{fig:exp}(b) presents the probability of detection versus different SNR ranges, for AWGN, single-path Rayleigh fading, and dual-path Rayleigh fading channels with delay spread $10\mu$s. In addition, Fig.~\ref{fig:exp}(b1) compares our results with the methods in \cite{yu2016} and \cite{qin2019}. 
As the figure shows, the proposed solution achieves more accurate results compared to previous works. For example, at $0$~dB SNR, probability of detection is between $70\%$ to $85\%$ which is higher than  \cite{yu2016} and \cite{qin2019}. 
The proposed method reaches about $100\%$ accuracy at about $2$~dB SNR while the other methods reach this point at about $4$~dB SNR. 


\subsection{Turbo Codes}
\label{sec:exp:3}

The candidate set for the turbo codes in our experiments consists of code rate $1/3$ with four different generator sequences as the following: $T1=(33, 25, 17)$, $T2=(15, 13, 24)$, $T3=(27, 37, 15)$, $T4=(25, 27, 37)$, and feedback sequence $33$, $37$, $35$, and $31$, respectively. Constraint length is $5$ and information length is $400$. 
Iterative turbo decoder algorithm is employed to decode the codes. 
Fig.~\ref{fig:exp}(c) presents the probability of detection versus different SNR ranges, for AWGN, single-path Rayleigh fading, and dual-path Rayleigh fading channels with delay spread $120\mu$s. 
The proposed method reaches more than $60\%$ accuracy after about $-5$~dB SNR, $-3$~dB SNR, and $5$~dB SNR under AWGN, fading and multi-path fading conditions, respectively.

\subsection{Polar Codes}
\label{sec:exp:5}

The candidate set for the polar codes in our experiments consists of four different rate-matched output length as the following: $P1=150$, $P2=160$, $P3=170$, and $P4=180$, with input length of $144$. 
Successive-cancellation list decoder of length 8 is used as decoders in the parallel decoders stage. 
Fig.~\ref{fig:exp}(d) presents the probability of detection versus different SNR ranges, for AWGN, single-path Rayleigh fading, and dual-path Rayleigh fading channels with delay spread $120\mu$s. 
The proposed method reaches more than $60\%$ accuracy after about $3$~dB SNR, $10$~dB SNR, and $13$~dB SNR under AWGN, fading and multi-path fading conditions, respectively. 


\section{Conclusion}
\label{sec:conc}

We considered the problem of recovering channel code parameters over a candidate set from the received encoded data. We proposed to decode the data with different parameters in the candidate set by common decoding algorithm and then feed the stack of them to convolutional neural network in order to recover the true code parameters. 
The proposed method works for several coding schemes and channel impairment settings. 
Experiments showed the proposed method outperforms previous works in different coding schemes. 
 

\begin{thebibliography}{10}
	\providecommand{\url}[1]{#1}
	\csname url@samestyle\endcsname
	\providecommand{\newblock}{\relax}
	\providecommand{\bibinfo}[2]{#2}
	\providecommand{\BIBentrySTDinterwordspacing}{\spaceskip=0pt\relax}
	\providecommand{\BIBentryALTinterwordstretchfactor}{4}
	\providecommand{\BIBentryALTinterwordspacing}{\spaceskip=\fontdimen2\font plus
		\BIBentryALTinterwordstretchfactor\fontdimen3\font minus
		\fontdimen4\font\relax}
	\providecommand{\BIBforeignlanguage}[2]{{%
			\expandafter\ifx\csname l@#1\endcsname\relax
			\typeout{** WARNING: IEEEtran.bst: No hyphenation pattern has been}%
			\typeout{** loaded for the language `#1'. Using the pattern for}%
			\typeout{** the default language instead.}%
			\else
			\language=\csname l@#1\endcsname
			\fi
			#2}}
	\providecommand{\BIBdecl}{\relax}
	\BIBdecl
	
	\bibitem{marazin2011}
	M.~Marazin, R.~Gautier, and G.~Burel, ``Blind recovery of k/n rate
	convolutional encoders in a noisy environment,'' \emph{EURASIP Journal on
		Wireless Communications and Networking}, p. 168, 2011.
	
	\bibitem{bonvard2018}
	A.~Bonvard \emph{et~al.}, ``Classification based on euclidean distance
	distribution for blind identification of error correcting codes in
	noncooperative contexts,'' \emph{IEEE Trans. on Signal Processing}, vol.~66,
	no.~10, pp. 2572--2583, 2018.
	
	\bibitem{xu2019}
	Y.~Xu, Y.~Zhong, and Z.~Huang, ``An improved blind recognition method of the
	convolutional interleaver parameters in a noisy channel,'' \emph{IEEE
		Access}, vol.~7, pp. 101\,775--101\,784, 2019.
	
	\bibitem{swaminathan2018}
	R.~Swaminathan \emph{et~al.}, ``Blind reconstruction of reed-solomon encoder
	and interleavers over noisy environment,'' \emph{IEEE Trans. on
		Broadcasting}, vol.~64, no.~4, pp. 830--845, 2018.
	
	\bibitem{shen2019}
	B.~Shen, H.~Wu, and C.~Huang, ``Blind recognition of channel codes via deep
	learning,'' in \emph{IEEE Global Conference on Signal and Information
		Processing}, 2019, pp. 1--5.
	
	\bibitem{wang2020}
	J.~Wang \emph{et~al.}, ``Fine-grained recognition of error correcting codes
	based on 1-d convolutional neural network,'' \emph{Digital Signal
		Processing}, vol.~99, p. 102668, 2020.
	
	\bibitem{ni2020}
	Y.~Ni \emph{et~al.}, ``Blind identification of ldpc code based on deep
	learning,'' in \emph{International Conference on Dependable Systems and Their
		Applications}, 2020, pp. 460--464.
	
	\bibitem{xia2014}
	T.~Xia and H.-C. Wu, ``Novel blind identification of ldpc codes using average
	llr of syndrome a posteriori probability,'' \emph{IEEE Trans. on Signal
		Processing}, vol.~62, no.~3, pp. 632--640, 2013.
	
	\bibitem{moosavi2014}
	R.~Moosavi and E.~G. Larsson, ``Fast blind recognition of channel codes,''
	\emph{IEEE Trans. on Communications}, vol.~62, no.~5, pp. 1393--1405, 2014.
	
	\bibitem{yu2016}
	P.~Yu, H.~Peng, and J.~Li, ``On blind recognition of channel codes within a
	candidate set,'' \emph{IEEE Comm. Letters}, vol.~20, no.~4, pp. 736--739,
	2016.
	
	\bibitem{condo2017}
	C.~Condo \emph{et~al.}, ``Blind detection with polar codes,'' \emph{IEEE Comm.
		Letters}, vol.~21, no.~12, pp. 2550--2553, 2017.
	
	\bibitem{wu2018}
	Z.~Wu, L.~Zhang, and Z.~Zhong, ``A maximum cosinoidal cost function method for
	parameter estimation of rsc turbo codes,'' \emph{IEEE Comm. Letters},
	vol.~23, no.~3, pp. 390--393, 2018.
	
	\bibitem{wu2019}
	Z.~Wu \emph{et~al.}, ``Blind recognition of ldpc codes over candidate set,''
	\emph{IEEE Comm. Letters}, 2019.
	
	\bibitem{sun2019}
	H.~Sun \emph{et~al.}, ``A novel blind detection scheme of polar codes,''
	\emph{IEEE Comm. Letters}, vol.~23, no.~8, pp. 1289--1292, 2019.
	
	\bibitem{qin2019}
	X.~Qin \emph{et~al.}, ``Deep learning based channel code recognition using
	textcnn,'' in \emph{International Symposium on Dynamic Spectrum Access
		Networks}, 2019, pp. 1--5.
	
	\bibitem{jalali2020}
	A.~Jalali and Z.~Ding, ``Joint detection and decoding of polar coded 5g control
	channels,'' \emph{IEEE Trans. on Wireless Communications}, 2020.
	
	\bibitem{condo2018}
	C.~Condo \emph{et~al.}, ``Design and implementation of a polar codes blind
	detection scheme,'' \emph{IEEE Trans. on Circuits and Systems II: Express
		Briefs}, vol.~66, no.~6, pp. 943--947, 2018.
	
	\bibitem{swaminathan2017(2)}
	R.~Swaminathan and A.~Madhukumar, ``Classification of error correcting codes
	and estimation of interleaver parameters in a noisy transmission
	environment,'' \emph{IEEE Trans. on Broadcasting}, vol.~63, no.~3, pp.
	463--478, 2017.
	
	\bibitem{xia2014(2)}
	T.~Xia, H.-C. Wu, and S.~Mukhopadhyay, ``Ldpc encoder identification in
	time-varying flat-fading channels,'' in \emph{IEEE Global Communications
		Conference}, 2014, pp. 3537--3542.
	
	\bibitem{liu2018}
	Y.~Liu \emph{et~al.}, ``Blind identification of ldpc codes in multipath fading
	channel via expectation maximization,'' in \emph{IEEE Global Communications
		Conference}, 2018, pp. 1--6.
	
	\bibitem{liu2020}
	Y.~Liu and F.~Wang, ``Blind data detection with unknown channel coding,''
	\emph{IEEE Comm. Letters}, 2020.
	
	\bibitem{wang2020nested}
	X.~Wang \emph{et~al.}, ``Nested construction of polar codes for blind
	detection,'' \emph{IEEE Wireless Communications Letters}, vol.~9, no.~5, pp.
	711--715, 2020.
	
	\bibitem{batch_norm}
	S.~Ioffe and C.~Szegedy, ``Batch normalization: Accelerating deep network
	training by reducing internal covariate shift,'' \emph{arXiv preprint
		arXiv:1502.03167}, 2015.
	
	\bibitem{dropout}
	N.~Srivastava \emph{et~al.}, ``Dropout: a simple way to prevent neural networks
	from overfitting,'' \emph{The Journal of Machine Learning Research}, vol.~15,
	no.~1, pp. 1929--1958, 2014.
	
	\bibitem{SourceCode}
	\BIBentryALTinterwordspacing
	Source code. [Online]. Available: \url{http://lis.ee.sharif.edu}
	\BIBentrySTDinterwordspacing
	
\end{thebibliography}


\end{document}